# Tunable mid-wave infrared Fabry-Perot bandpass filters using phase-change GeSbTe

Calum Williams[1], Nina Hong[2], Matthew Julian[3,4], Stephen Borg[5], Hyun Jung Kim[4,5*]

[1.] *Department of Physics, Cavendish Laboratory, University of Cambridge, Cambridge, CB3 0HE, UK*
[2.] *J.A. Woollam Co., Inc, Lincoln, NE 68508, USA*
[3.] *Charles L. Brown Department of Electrical and Computer Engineering, University of Virginia, VA, 22904, USA*
[4.] *National Institute of Aerospace, Hampton VA 23666, USA*
[5.] *NASA Langley Research Center, Hampton VA 23681-2199, USA*

\* *hyunjung.kim@nasa.gov*

## Abstract

We demonstrate spectrally-tunable Fabry-Perot bandpass filters operating across the MWIR by utilizing the phase-change material GeSbTe (GST) as a tunable cavity medium between two (Ge:Si) distributed Bragg reflectors. The induced refractive index modulation of GST increases the cavity's optical path length, red-shifting the passband. Our filters have spectral-tunability of ~300 nm, transmission efficiencies of 60-75% and narrowband FWHMs of 50-65 nm (Q-factor ~70-90). We further show multispectral thermal imaging and gas sensing. By matching the filter's initial passband to a $CO_2$ vibrational-absorption mode (~4.25 µm), tunable atmospheric $CO_2$ sensing and dynamic plume visualization of added $CO_2$ is realized.

# 1. Introduction

The mid-wave infrared (MWIR) waveband (~3–5 µm) is invaluable for sensing and imaging systems, from probing molecular vibrations in chemical species to detecting radiant thermal signatures [1–4]. MWIR imaging devices play a key role in many diverse technological applications, including multi/hyperspectral imaging, thermography, chemical spectroscopy, surveillance, automotive safety, and astronomy [5–8]. Tunable optical properties — such as sampling more than single wavelength and operating across wide wavelength band — are highly desirable for maximizing the available detectable information [9][10]. Yet achieving this in a single compact solid-state device remains challenging, with no single technology providing a complete solution.

The majority of wide-field visible-MWIR imaging systems contain passive Fabry-Perot (FP)-based bandpass filters comprising multi-layer dielectric, or metal-dielectric, thin-film stacks, either spatially encompassing the entire image sensor /focal plane array [5,11] or within multispectral filter array arrangements [10,12]. In its simplest form, a FP-bandpass filter consists of two distributed Bragg reflectors (DBRs) separated by a resonant single-cavity (spacer). The DBRs, acting as dielectric-mirrors, open up a stopband, with the resonant cavity providing a passband with narrow full-width-half-maximum (FWHM); layer thicknesses for each are designed for the centre wavelength (CWL) of operation [11,13]. Thin-film FP-bandpass filter design is well established [11,13] — with mass-produced filters still offering unrivalled optical performance when compared to alternative spectral filter technologies. However, once manufactured, these FP-bandpass filters typically offer no straightforward means to actively tune their spectral properties.

For 'active' spectral tunability, two operating mechanisms are generally utilized: (1) modifying the physical-optical properties of constituent material/s, or (2) mechanically altering the device design /system setup [14–16]. For the former, the optical properties of materials can be tuned by: (a) employing materials with adjustable refractive indices (i.e. index modulation capability), and or (b) incorporating patterns (i.e. gratings, waveguides etc.), then changing geometry (size, shape etc.). For (a), liquid crystals (LCs) — with voltage controllable birefringence properties — have been widely implemented (e.g. LC-Lyot filters) to provide spectral filtering [17,18]. However, this costly approach is inherently polarization-dependent, offers low transmission efficiencies, operates primarily in the visible — as LCs typically exhibit vibrational-absorption modes in the IR — and provides relatively slow switching speeds (~kHz) [18,19]. Another related technology are acousto-optic tunable filters (AOTFs), whereby the optical properties of a birefringent crystal, and subsequent diffraction output, can be controlled through acoustic waves generated via applied RF signals [18,20]. This approach, albeit fast-switching (~MHz), is costly, requires bulky hardware, has small entrance apertures and hence is generally unsuitable for wide-field imaging purposes. Over recent years, plasmonic /all-dielectric nanostructure arrays have been developed for compact, ultra-thin spectral filters [21–23]. These devices provide spectral filtering through the excitation of electric /magnetic resonances in tailored metallic or high-index dielectric nanostructure arrays. Nonetheless, for bandpass filters, their transmission response is typically broad with low efficiencies; polarization-dependent multipolar (additional) modes are often excited in the same spectral region, and ultra-high resolution lithographic techniques (e.g. deep /extreme UV) are required for commercial adoption [21–23]. For active tunability, integration with electro-optic materials such as LCs [24–26] is commonly used.

For 'real-world' wide-field multispectral imaging applications — such as remote sensing and thermal (IR) imaging — motorized filter wheels are widely utilized [9,27,28]. These systems contain many passive narrowband filters in a circular array, and through mechanical tuning (stepped rotation), the filters are time-sequentially positioned in the optical beam path. This approach has slow response times, provides limited spectral resolution (dependent on the number of the filters), is bulky, and has moving parts. A single compact solid-state tunable filter which is fast switching, has no moving parts, is straightforward to operate and inexpensive to manufacture is highly desirable for many fields underpinned by IR-spectroscopic methods and for filter wheel replacement.

Recently, phase change materials (PCMs) have gained interest as a new platform for tunable optical devices due to their pronounced refractive index contrast (between disordered-amorphous and ordered-crystalline states), fast switching speeds, and good thermal stability [29–31]. The prototypical chalcogenide PCM, $Ge_2Sb_2Te_5$ (GST), is non-volatile, can be reversibly switched on a nanosecond timescale and exhibits large index modulation (~2.4) across the IR [29–32]. When GST is heated above its glass transition temperature — through laser pulse or electronic excitation — it produces a thermal transition occurring through nucleation and crystallization; 150ºC for face-centred cubic (FCC) packing, 360ºC for hexagonal close packed (HCP) growth [29–31,33]. Re-amorphization is achieved by heating the material above its melting temperature (632ºC) followed by quenching. This optical contrast is a key property of PCMs for its widespread commercial application in optically rewritable data storage device and increasingly common usage in tunable /reconfigurable micro-optical devices such as waveguides, variable-focal lenses and filters [30,31,33–36]. PCMs maintain their structural state and only require energy during the switching process, which is a clear advantage over liquid crystals and mechanically tuned photonic devices. Moreover, GST is both cost-effective and scalable for large-area integration and exhibits tunable optical properties across broad wavebands.

In this work, we utilize GST as a phase-change cavity embedded between Ge:Si DBRs to enable spectrally-tunable (bi-stable) all solid-state FP-bandpass filters operating across the MWIR waveband. The optical path length of the GST cavity, hence resultant passband CWL, is spectrally red-shifted under phase transformation due to GSTs induced refractive index modulation. Therefore, through external stimuli, our filter's CWL can be spectrally switched to two different states (passbands); a reversible process. Our 1-inch diameter filters have narrowband FWHMs of 50–65 nm (with Q-factor ~70–90) and high transmission efficiencies of 60–75%. We further show multispectral thermal (IR) imaging capability by integrating our filters with a commercial thermal camera and imaging two separate wavelengths. Moreover, by designing the GST-BP filter's CWL to correspond to a molecular vibration mode of $CO_2$ (~4.25 μm), we demonstrate gas sensing (imaging) of both atmospheric (native) $CO_2$ and externally added CO2 gas; imaging the dynamic plume for the latter. Our GST-based thin-film filters have wide applicability for IR imaging and sensing applications which require compact, cost-effective tunable spectral filters.

## 2. Device Concept

The simplest thin-film optical bandpass filters are composed of a single-spacer (resonant cavity) between two multi-layer dielectric mirrors [11,13,37]. The dielectric mirrors — commonly known as distributed bragg reflectors (DBRs) — consist of a series of alternating high-and-low index materials (i.e. N bi-layers) with ¼-wave optical thickness, which provides a spectral 'rejection region' [11,13,37]. An embedded spacer opens up a transmission passband, with optical thickness governing the filter's center wavelength (CWL) of operation. A material's optical thickness is dependent on its refractive index, hence by utilizing the phase-change material GST — which exhibits a significant index modulation across the waveband of interest (MWIR) — as the spacer material, it is possible to control the passband CWL through the phase transformation of GST.

As our bandpass filters are designed for operation across the MWIR, Ge (H – *high index*) and Si (L – *low index*) are chosen as the DBR materials, with $CaF_2$ as the substrate. GST is then utilized as a tunable spacer material, with capability to operate in either its amorphous (a-GST) or crystalline (c-GST) state (induced through external stimuli). Our tunable resonant cavity design concept is shown schematically in Fig 1a. The simulated transmission response of this design, for the two GST states, is shown in Fig 1b as a function of N bi-layers. From this, N = 6 bi-layers were chosen for the proof of concept filters — a compromise between the filters' optical properties and total deposition time — with simulated response of this arrangement shown in Fig 1c. We expect a resultant passband spectral shift of ~300nm between the two GST states along with minimal decrease in transmission and small increase in FWHM.

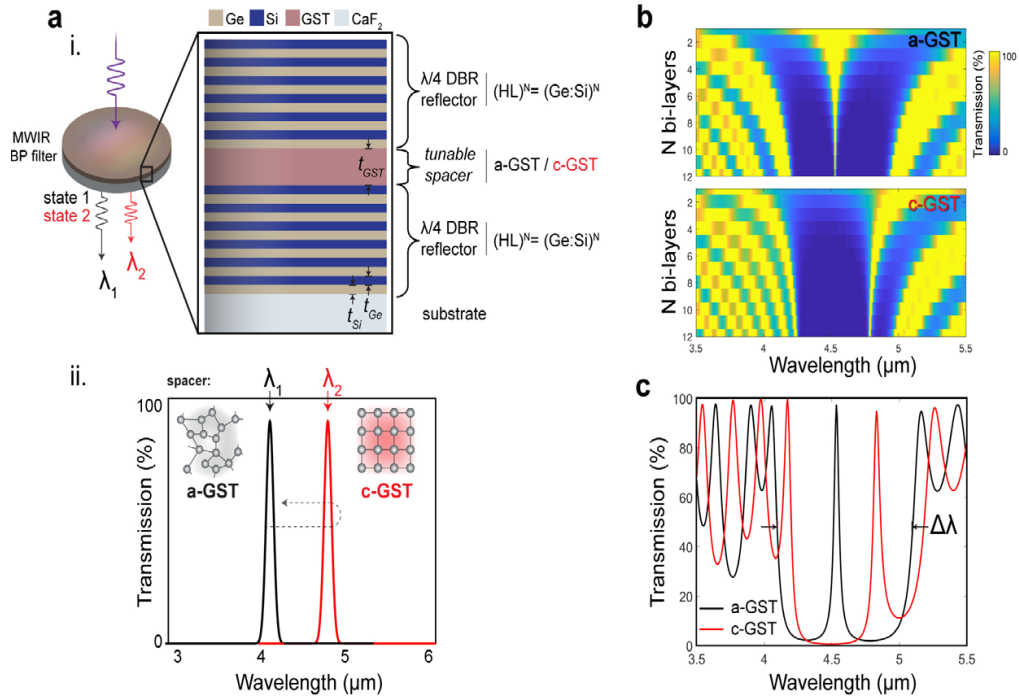

**Fig. 1. Device concept and simulations.** (a) Tunable MWIR Fabry-Perot bandpass filter layer design, with thicknesses of filter materials (e.g. $t_{GST}$) indicated (i), whereby the centre wavelength ($\lambda_1$ or $\lambda_2$) of the narrowband transmission response will spectrally shift (ii) (a reversible process) depending on the spacer's optical thickness, hence GST crystallinity (refractive index). (b) Simulation of the effect of N-bi layers on the transmission response of the filter for a-GST and c-GST spacer layers with quarter-wave DBR stacks. (c) Simulation of the 6-layer designed FP filter. DBR = distributed Bragg reflector, BP = bandpass, HL = high-low index bi-layer, $\Delta\lambda$ = blocking wavelength range (rejection region).

# 3. Materials and Methods

*3.1 GST-FP bandpass filter: thin film materials deposition and characterization*

Double-side polished 1.5 x 25.4 mm (1-inch optics) $CaF_2$ wafers (*Esco Optics, Inc*) were sonicated in acetone, isopropanol, and deionized water sequentially, twice, followed by ultra-high purity compressed nitrogen ($N_2$) blow dry, before deposition. All thin-films were deposited via 50W RF magnetron sputtering at a base pressure of 2.6 x $10^{-6}$ Torr and a deposition pressure of 7 mTorr (10 sccm Ar flow, research-grade, 99.9999% purity). Three separate sputtering targets were used for the high-index, low-index and cavity (spacer) layers: Ge (99.999% purity), Si (99.999% purity) and $Ge_2Sb_2Te_5$ target (14.3 wt% Ge, 23.8 wt% Sb, 61.9 wt% Te, *Mitsubishi* materials, Inc.). The chemical composition of as-deposited GST thin-films was determined by DCP-AES (Direct Current Plasma-Atomic Emission Spectroscopy, *Luvak, Inc.*). The composition was measured as 22 at% Ge, 23.5 at% Sb, and 54.5 at% Te, which is close to a nominal composition of bulk $Ge_{22.2}Sb_{22.2}Te_{55.6}$ [33]. Fourier Transform Infrared (FTIR) spectroscopy (*ThermoFisher*) was used to measure the transmission intensity data of the bandpass filters. X-ray diffraction (XRD) was used to determine the crystallinity, and ellipsometry used to measure complex complex refractive indices of the respective layers; RC2 ellipsometry system (Model: DI, *J.A. Woollam*) to measure up to 1.6 µm and IR-VASE (Infrared Variable-angle spectroscopic ellipsometry, *J.A. Woollam*) to measure up to 7 µm. Results of both the XRD and ellipsometry measurements of as-deposited a-GST and c-GST (HCP) are shown in Fig 2a,b and Fig 2c,d respectively. A significant index modulation across the IR can be observed, with $\Delta n \sim 2.4$ at $\lambda = 4.5$ µm. Fig e,f shows the refractive index measurements for RF sputtered thin-film Ge and Si (high-low bi-layer respectively). The measured refractive index values are summarized in Table 1 and compare closely with similar studies in the literature [29–31]. These values are subsequently used to determine the ¼-wave thicknesses for the fabricated FP-bandpass filters.

*3.2 Multi-layer stack numerical simulation and layer thicknesses*

Recursive propagation matrices (transfer matrix method, implemented in MATLAB [13]) are employed to determine the optical output of the thin-film multilayer dielectric stacks [11,13]. The transmission and reflection coefficients at each interface of the thin-film stack are calculated as a function of wavelength, and overall transmission response determined recursively [13]. Each layer is assigned its material-specific refractive index, with values obtained experimentally through ellipsometry measurements. For the two DBRs either side of the GST spacer, N = 6 bi-layers with thicknesses = $\lambda_0/4n$, where $n$ is the experimentally determined refractive index of each high-low layer and $\lambda_0$ the design CWL. As such, the stack arrangement is $(HL)^N(spacer)(HL)^N = (Ge:Si)^6(GST)(Ge:Si)^6$, where GST can be in either its amorphous or crystalline state. At 4.5 µm, an index modulation of $(\Delta n) = 2.4$ exists between the two GST states (Table 1) which provides the change in optical path length for the spacer and thus the spectral shift in passband CWL. The comparably low number of bi-layers (with respect to commercial BP-filters) was chosen as a balance between desirable filter performance (narrow FWHM, high transmission etc.) and total deposition /processing time for a proof-of-concept set of devices.

*3.3 Infrared imaging demonstrations*

A hotplate, with a NASA insignia logo (milled from a 15 cm x 15 cm x 3 mm aluminum plate) placed in front, was used to mimic a spatially-variant blackbody thermal source. The hotplate (thermal source) was heated from 320K to 486K and an IR camera (FLIR SC8300HD, 0.02 $Wm^{-2}$ minimum pixel sensitivity) connected to a PC, was used to capture the images. A field stop was used in order to eliminate unwanted signal from polluting the measurement data. Our GST-based FP bandpass filters were mounted and positioned between the IR camera and thermal source.

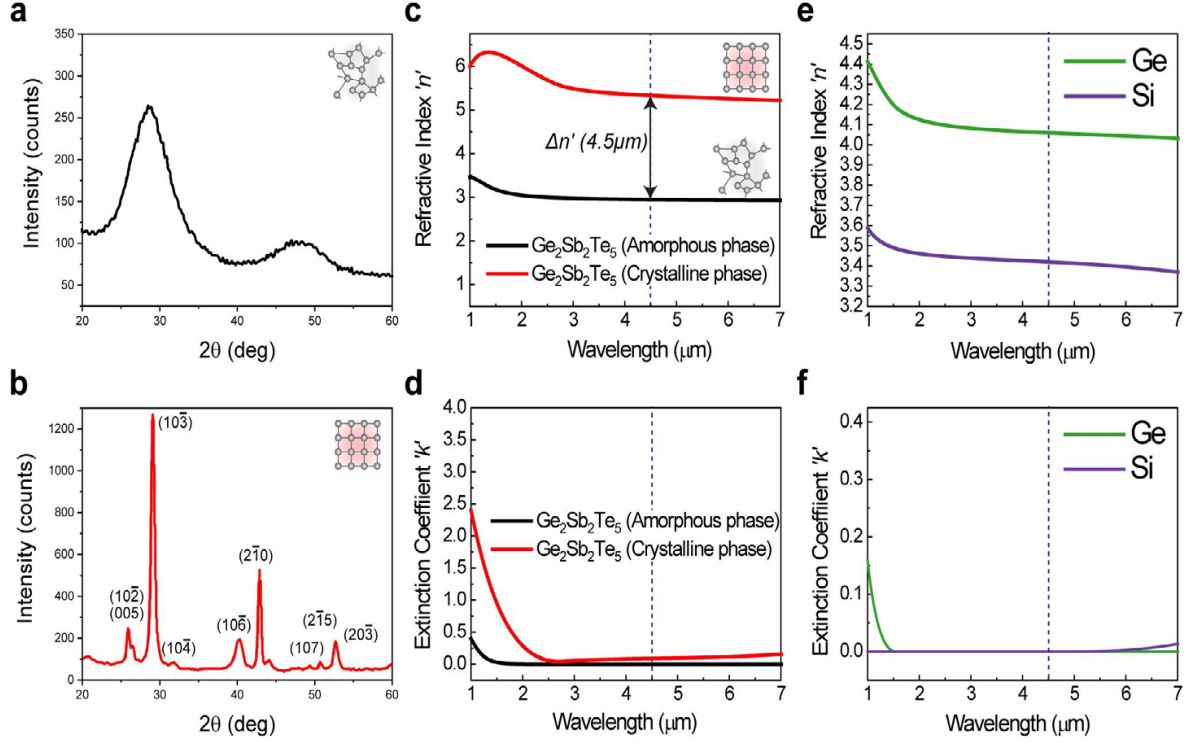

**Fig. 2. Thin-film materials characterization.** XRD data for sputtered amorphous $Ge_2Sb_2Te_5$ (a) and (b) crystalline (HCP) $Ge_2Sb_2Te_5$ states. (c) Real and imaginary (d) part of the refractive index for thin-film a-GST and c-GST across $\lambda = 1\text{-}7\mu m$, obtained through ellipsometry, with index modulation ($\Delta n'$) of ~2.4 at 4.5μm indicated. (e) Real and imaginary (f) part of the refractive index for sputtered Ge and Si across $\lambda = 1\text{-}7\mu m$, obtained through ellipsometry. The design CWL of 4.5μm is overlaid for visual reference.

**Table 1.** Summary of the experimentally derived MWIR optical constants (at 4.5μm and 4.52μm) of the RF sputtered thin-films: Si, Ge, and $Ge_2Sb_2Te_5$ (GST-225) in its amorphous and crystalline states

| Material | $\lambda = 4.5\mu m$ | | $\lambda = 4.2\ \mu m$ | |
|---|---|---|---|---|
| | n | k | n | k |
| Ge | 4.100 | ~0 | 4.103 | ~0 |
| GST-225 (*amorphous*) | 2.948 | 0 | 2.952 | ~0 |
| GST-225 (*crystalline*) | 5.336 | 0.089 | 5.354 | 0.086 |
| Si | 3.42 | ~0 | 3.424 | ~0 |

## 4. Tunable GST-FP bandpass filter operating in the MWIR

Two bandpass filters (25.4mm diameter) with N = 1 and N = 6 alternating (HL)$^N$ bi-layers, and layer thicknesses designed for $\lambda_0$ = 4.5 μm, were fabricated and optically characterized (Fig 3a). With the arrangement (Ge:Si)$^N$(GST)(Ge:Si)$^N$, the physical layer thicknesses were, $t_{Si}$ = 329 nm, $t_{Ge}$ = 280 nm and $t_{GST}$ = 375 nm spacer (single cavity). The filters were annealed at 400°C inside a vacuum chamber to induce GST crystallization, then re-measured. Fig 3b shows the transmission results of the N = 1 GST-BP filter for the two GST states. The a-GST state has CWL ~ 4.5 μm with 85% transmission efficiency, and c-GST state CWL ~ 4.9 μm with 70% transmission efficiency. The longer wavelength state (with c-GST spacer) has a slightly reduced transmission peak attributed to the increased infrared absorption (extinction coefficient, Fig 2d) of the c-GST compared to a-GST. Fig 3c compares the transmission spectra of the two (N = 1 and N = 6) BP filters with the initial a-GST spacer. As expected, the FWHM (~ 50 nm) narrows at the CWL, but the transmission decreases, which in this case we speculate is attributed to cumulative optical scattering from grain boundary defects of the multiple layers. The blocking wavelength range ($\Delta\lambda$) is ~ 4.1–5 μm, and higher order 'ripples' can be observed outside this region. Fig 3d compares the N = 6 filter in its two operating states. The initial a-GST state has CWL ~ 4.5 μm (as designed) with 75% transmission efficiency, and c-GST state CWL ~ 4.75 μm with 60% transmission efficiency and slight increase in FWHM to ~65 nm (corresponding to a Q-factor ~ 70–90). This proof-of-concept filter design can be further improved through incorporation of multiple-cavity GST-cavities and optimization of additional layers which will suppress out-of-band ripples [11,38,39].

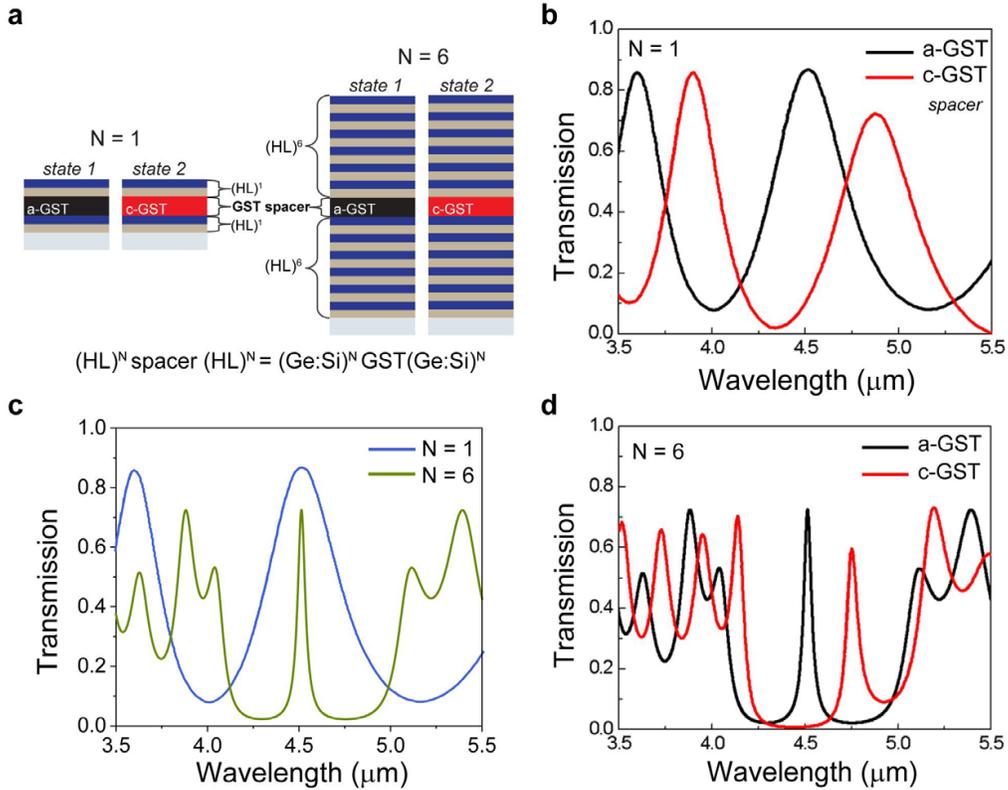

**Fig. 3**. **Tunable GST-FP MWIR bandpass filter characterization.** (a) Schematic of the two BP filters, with N=1 and N=6 bilayers, which can operate in two states: a-GST or c-GST spacer. (b) Transmission spectra of the two GST spacer states for a N=1 bi-layer stack - (Ge:Si)$^N$(GST)(Ge:Si)$^N$. (c) Transmission spectra of the two filter types (with N=1 and N=6) in their a-GST states. (d) Transmission spectra of the N=6 filter in its two states: a-GST and c-GST.

## 5. Applications of GST-FP filters in the MWIR

*5.1 Thermal imaging demonstration*

Our GST-FP bandpass filter design (Fig 1d), with CWL(a-GST) ~ 4.5 µm and CWL(c-GST) ~ 4.75 µm, is now duplicated with each set to one of the two GST crystallinities. i.e. filter 1: a-GST spacer, filter 2: c-GST spacer. A hotplate, with Al NASA Insignia logo transmission mask placed in front, is used as a spatially varying thermal-IR object (blackbody source). Between 320K to 486K, the blackbody emissive spectral peak power blue shifts; from ~9 µm to 5.8 µm, with total integrated power increasing in the MWIR band [33,40]. The emissive power is higher at 4.75 µm than 4.5 µm, and this difference increases as the blackbody temperature increases. Thus, a difference in transmission intensity (counts) between the two filters, which increases as hotplate temperature increases, is expected.

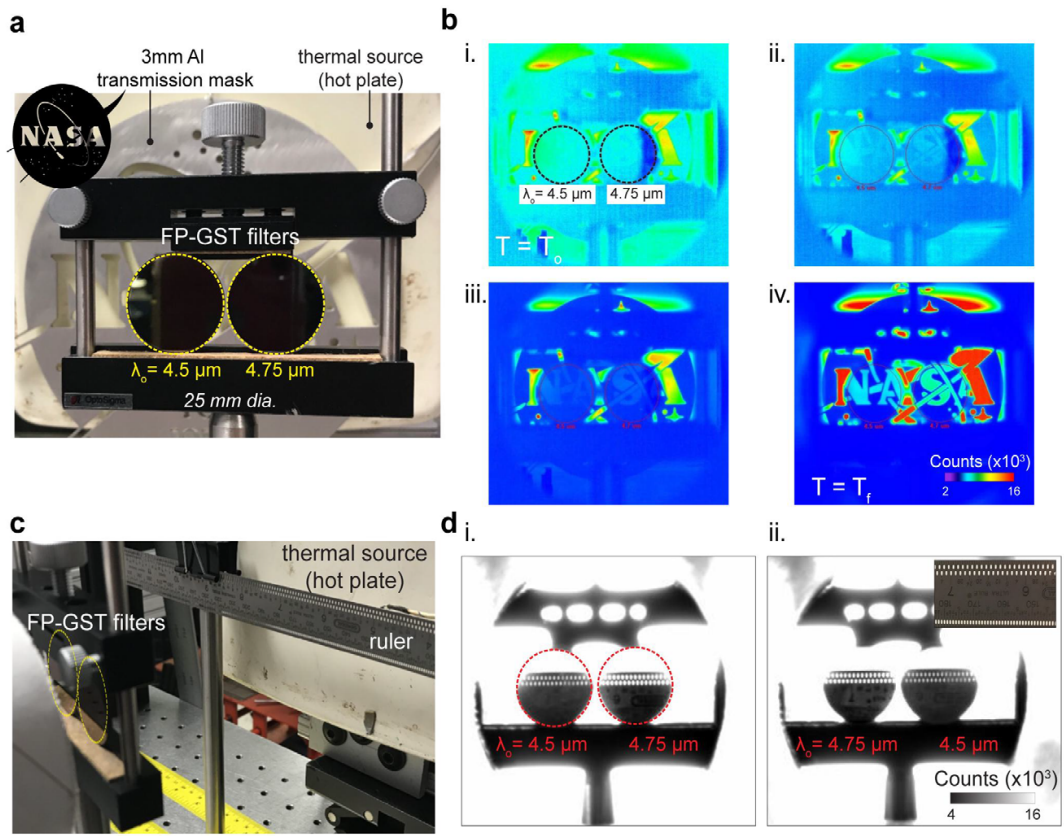

**Fig. 4. Thermal imaging with GST-FP bandpass filters**. (a) White light (RGB) image of the thermal imaging setup showing the hotplate, NASA Insignia logo (transmission mask) and the two mounted filters (*left*) a-GST spacer, and (*right*) c-GST spacer, with CWL(a-GST) ~ 4.5 µm and CWL(c-GST) ~ 4.75 µm. (b) Shows the thermal (IR) image of the same scene but using a MWIR digital camera. The hotplate temperature (blackbody source) is increased from 320K (i) to 486K (iv). (c,d) Comparison of the 4.5 µm and 4.7 µm filter at 515K. (a) test setup, (b) the smaller objects (i.e. higher spatial frequencies) image from 4.5 µm than the 4.7 µm, and (c) the filter positions has been swapped to check repeatability.

The two filters are placed between the thermal object (hotplate) and a MWIR digital camera. This arrangement is shown under white light (RGB) conditions in Fig 4a. Fig 4b shows the MWIR camera image data as the hotplate increases in base temperature, from $T_0$ = 320K (i) to $T_f$ = 486K (iv). A variable transmission response through the two filters, and subsequent identification of the NASA Insignia logo (spatially variant thermal profile), can be observed. As the temperature increases, the intensity captured through the longer wavelength filter is greater than that of the shorter wavelength filter. Note, due to the increased extinction coefficient for the c-GST state, thus longer wavelength filter, the difference in intensity is slightly less than expected if the two filters had the same CWL transmission efficiency. The performance of our filters is further demonstrated through the imaging (resolving) of a metallic ruler (Fig 4. c,d) placed between source (set at 486K) and camera. It can be observed that the intensity is higher for the longer wavelength filter, the ruled lines (and holes) can be resolved and that the filters' intensity response is spatially uniform over the entire 1-inch area.

*5.2 Carbon dioxide gas sensing*

The transmittance of electromagnetic radiation through the atmosphere is dependent on the relative amount of constituent gases, hence amounts of radiation-absorbing molecules [5,41]. Carbon dioxide ($CO_2$) for example, has two molecular (vibrational) absorption modes in the MWIR: 2.6–2.7 µm and ~4.2–4.3 µm (Fig 5a) [41], with large (~10:1) transmission modulation at the longer wavelength mode. By designing our GST-FP bandpass filter to have its initial (a-GST) CWL centred at approximately the longer wavelength $CO_2$ absorption peak, an induced phase-change would red-shift the filter's CWL (c-GST state) to beyond the absorption peak, hence 'seeing through' (filtering out) the $CO_2$ information and transmitting at higher intensities. Fig. 5b shows the experimentally measured transmission spectra of this $CO_2$ GST-FP filter. With N (bi-layer) = 6 arrangement as before, the physical layer thicknesses are now, $t_{Si}$ = 311 nm, $t_{Ge}$ = 260 nm and $t_{GST}$ = 355 nm spacer (cavity). Again, two filters are fabricated, one set in its initial (a-GST) state, with CWL ($\lambda_1$ ~4.26 µm), and the other annealed to its secondary (c-GST) state, with CWL ($\lambda_2$ ~4.55 µm). They are positioned in front of the hotplate (set at T = 486K) and a supply of pure $CO_2$ gas is fed, via small inlet nozzle, between the hotplate and the filters; setup shown in Fig 5c. The $CO_2$ gas supply is varied and MWIR camera is used to capture the resultant response (Fig 5d and *Supplementary Video 1*). Initially, the gas line valve is closed, with no $CO_2$ gas flowing from the inlet, hence only $CO_2$ within the atmosphere is present (Fig 5d, i). A uniform transmission intensity (counts) difference between both filters can be observed, with the 4.26 µm filter much lower in intensity as expected. The $CO_2$ gas flow is then varied (Fig 5d, ii-iv) and the real-time $CO_2$ gas flow pattern is captured through the 4.26 µm filter (Fig 5d and Supplementary Video 1) in the form of lower transmission spatially varying dynamic gas plume. The additional intensity contrast between background atmospheric (native) $CO_2$ and the added $CO_2$ can be observed. Even though two separate filters have been used for demonstration purposes here, clearly a single, switchable, GST-FP tunable bandpass filter of this design has the ability to both detect and resolve the variation in concentration of $CO_2$ gas.

Short term applications for these filters include both ground-based and remote-sensing based differential absorption LIDAR (DIAL). Such systems are critical for measuring the molecular composition of the atmosphere in order to better understand climate behavior and atmospheric contaminants, among other things. DIAL systems output two /several NIR- or IR-laser wavelengths corresponding to the 'on /off' resonance of the molecular absorption signature of specific greenhouse gases such as $CO_2$ in the MWIR, as well as other chemical species such as water vapor in the NIR [42,43]. This process requires ultra-fast switching of the laser wavelength to rapidly determine gaseous absorption contrast in the atmosphere and typically utilizes several optical sub-systems (including FP-etalons and acoustic-optical-modulators) for spectral filtering. Our GST-FP filters can be designed to operate at specific points across IR wavebands, due to both GSTs broadband index modulation at these wavelengths and through layer thickness design. In addition, GST can be switched on nanosecond timescales [29,42-43]. As a result, our cost-effective tunable GST-FP filters may be the ideal candidate for integration into DIAL sub-systems.

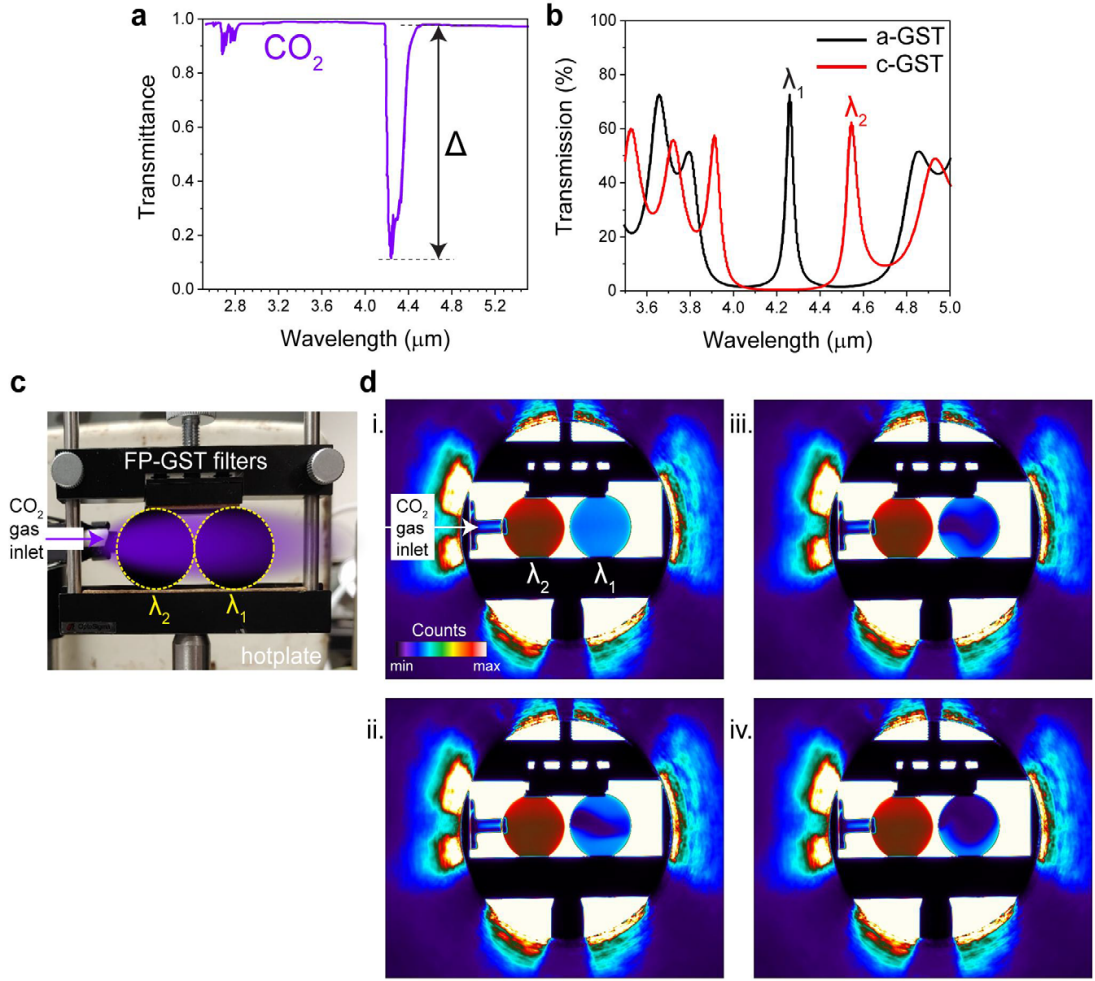

**Fig. 5. $CO_2$ gas sensing with GST-FP tunable bandpass filters.** IR transmittance spectra of $CO_2$ (gas phase) — data obtained from [38] — with intensity modulation ($\Delta$) between 4.25 μm and 4.55 μm indicated. (d) Transmission spectra of the two $CO_2$ gas filer states: a-GST and c-GST, which were used for imaging. (c) Image (RGB) of the gas sensing setup with two filters at the two respective CWLs ($\lambda_1$ and $\lambda_1$ indicated). (d) IR-imaging of the two GST-FP bandpass filters in front of the hotplate (set at T = 486K) with a supply of $CO_2$ gas fed via inlet nozzle, between the hotplate and the filters. The $CO_2$ gas supply, initially turned off (i), is varied and MWIR camera is used to capture the resultant response. The gas plume can be observed (ii-iv) as the supply is varied. Real-time capture video see *Supplementary Video 1*.

# 6. Conclusion

The MWIR (3-5 µm) waveband is of significant interest across different sensing and imaging applications, from FTIR chemical spectroscopy to remote sensing. Tunable all-solid state spectral filtering is highly desirable for maximizing the detectable spatial-spectral information. Widely utilized optical bandpass filters based on multi-layer Fabry-Perot thin-films still provide 'best-in-class' narrowband and high transmission performance in comparison to alternative methodologies such as LC-based or plasmonic /nanophotonic approaches. Nonetheless, actively tunable all solid-state spectral filtering in a single compact device still remains challenging, with bulky filter wheels and expensive MEMS-based devices providing sub-optimal (mechanical) solutions.

Here, we utilize phase-change alloy GeSbTe (GST) thin-films which enable tunable all solid-state FP-bandpass filters operating across the MWIR waveband. GST—a cost-effective, readily available alloy—is used as a tunable single-cavity (spacer) between two (Ge:Si)[6] DBR mirrors. Because GST exhibits a significant refractive index modulation (~2.4) in the MWIR between its amorphous and crystalline states, the optical path length of the cavity, hence resultant passband, is spectrally shifted under phase transformation. Therefore, through external stimuli, our filter's CWL is spectrally switched; a reversible process. We experimentally demonstrate two tunable bandpass filter designs: the first, with CWLs ~4.5 µm and 4.75 µm, and the second; CWLs ~4.25 µm and 4.55 µm. The filters — which are 1-inch in diameter — have FWHMs of 50–65 nm (with Q-factor ~70–90) and transmission efficiencies of 60–75%. We further show 'real-world' applicability by performing multispectral thermal (IR) imaging with our filters in front of a commercial wideband MWIR camera. A NASA Insignia logo, in front of a blackbody thermal source (hotplate) is imaged through our filters. In addition, the second filter is designed to have its initial CWL (amorphous GST state) matched to a molecular (vibrational) absorption mode of $CO_2$ and demonstrate gas sensing (imaging) of both atmospheric (native) $CO_2$ and externally added $CO_2$ gas. For the latter, the dynamic gas plumes are captured through the intensity contrast between atmospheric and added gas, indicating that the filter has the ability to both detect and resolve the variation in concentration of $CO_2$ gas.

Filter operation is purely bi-stable in this paper — with the GST-spacer in either is fully-amorphous or fully-crystalline states — yet continuous CWL tuning across the stopband should be possible by utilizing the partial-crystallinity states of GST [29,33]. These intermediate phase states (i.e. a-GST, to $p_1$-GST...$p_{final}$-GST, to c-GST) result in a gradual change in refractive index, hence smaller change in optical path length of the cavity, thus filter CWL spectrally shifts in finer increments. To induce these states, the fine control /variation of external stimuli such as laser-induced energy densities is required [29,33].

Our tunable all solid-state GST-FP bandpass filters (which can be modified to operate across other wavebands) are straightforward to manufacture—requiring no lithographic step and use cheap materials—and provide narrowband high transmission optical performance. Therefore, they have wide applicability for imaging and sensing applications requiring compact cost-effective tunable filtering, which include DIAL and CubeSat based LIDAR optical sub-systems.

## Funding

H.J.K and S.B acknowledge support from the NASA LaRC CIF/IRAD Program. C.W. acknowledges support from the Cancer Research UK Pioneer Award (C55962/A24669), Engineering and Physical Sciences Research Council (EP/R003599/1) and through the Wellcome Trust Interdisciplinary Fellowship.

## Acknowledgments

The authors appreciate Mr. Joel Alexa at NASA LaRC with XRD analysis. H.J.K. would like to acknowledge the support of Mr. Thomas Jones and Mr. William Humphreys at NASA LaRC. C.W. would like to acknowledge the support of Dr. Sarah Bohndiek and Dr. Geroge Gordon.

## Disclosures

The authors declare no conflicts of interest.